\documentclass[runningheads]{llncs}

\usepackage{graphicx}
\usepackage{hyperref}
\usepackage{xcolor}
\usepackage{textcomp}
\usepackage{bbding}
\usepackage{amsmath}
\usepackage[ruled,vlined,linesnumbered]{algorithm2e}
\usepackage{subcaption}
\usepackage{booktabs}
\usepackage[font=scriptsize,labelfont=bf,tableposition=top]{caption}
\usepackage{multirow}
\SetKw{Continue}{continue}
\SetKw{Break}{break}

\hypersetup{
	colorlinks,
	linkcolor={red!50!black},
	citecolor={blue!60!black},
	urlcolor={blue!80!black}
}

\begin{document}

\title{Efficient Construction of Behavior Graphs\\ for Uncertain Event Data\thanks{In \emph{International Conference on Business Information Systems} (BIS 2020). We thank the Alexander von Humboldt (AvH) Stiftung for supporting our research interactions. Please do not print this document unless strictly necessary.}}

\author{Marco Pegoraro\orcidID{0000-0002-8997-7517} \and Merih Seran Uysal\orcidID{0000-0003-1115-6601} \and Wil M.P. van der Aalst\orcidID{0000-0002-0955-6940}}

\authorrunning{Pegoraro et al.}

\institute{Process and Data Science Group (PADS) \\ Department of Computer Science, RWTH Aachen University, Aachen, Germany
	\email{\{pegoraro, uysal, wvdaalst\}@pads.rwth-aachen.de}\\
	\url{http://www.pads.rwth-aachen.de/}}

\maketitle

\begin{abstract}
The discipline of process mining deals with analyzing execution data of operational processes, extracting models from event data, checking the conformance between event data and normative models, and enhancing all aspects of processes. Recently, new techniques have been developed to analyze event data containing uncertainty; these techniques strongly rely on representing uncertain event data through graph-based models capturing uncertainty. In this paper we present a novel approach to efficiently compute a graph representation of the behavior contained in an uncertain process trace. We present our new algorithm, analyze its time complexity, and report experimental results showing order-of-magnitude performance improvements for behavior graph construction.

\end{abstract}

\section{Introduction}
Process mining~\cite{van2016process} is a research field that performs process analysis in a data-driven fashion. Process mining analyses are based on recordings of events and tasks within the process, stored in a number of information systems supporting business activities. These recordings are extracted and orderly collected in databases called \emph{event logs}. Utilizing an event log as a starting point, process mining analyses can automatically extract a process model describing the behavior of the real-world process (\emph{process discovery}) and measure deviations between execution data of the process and a normative model (\emph{conformance checking}). Process mining is a rapidly growing field both in academia and industry. More than 25 commercial tools are available for analyzing processes. Process mining tools are used to analyze processes in tens of thousands of organizations, e.g., within Siemens, over 6000 employees use process mining to improve processes.

Commercial process mining tools are able to automatically discover and draw a process model from an event log. Most of the process discovery algorithms used by these tools are based on counting the number of \emph{directly-follows relationships} between activities in the process. The more often a specific activity follows another one in a process of an organization, the stronger a causality implication between the two activities is assumed to be. Directly-follows relationship are also the basis for detecting more complicated constructs in the workflow of a process, such as parallelism or interleaving of activities. These relationships are often summarized in a labeled graph called the \emph{Directly-Follows Graph} (DFG).

Recently, a new class of event logs has gained interest: \emph{uncertain event logs}~\cite{pegoraro2019mining}. These execution logs contain, rather than precise values, an indication of the possible values acquired by event attributes. In this paper, we will consider the setting where uncertainty is expressed by either a set or an interval of possible values for an attribute, as well as the possibility of an event being recorded in the log even though it did not occur in reality. An example of an uncertain trace is shown in Table~\ref{table:uncertaintrace}.

\begin{table}[]
	\caption{An example of simple uncertain trace. Events $e_2$ and $e_4$ have uncertain activity labels. Event $e_3$ has a possible range of timestamps, rather than a precise value. Event $e_5$ has been recorded, but it might not have happened in reality.}
	\label{table:uncertaintrace}
	\centering
	\begin{tabular}{ccccc}
		\textbf{Case ID} & \textbf{Event ID}        & \textbf{Activity}                                                                                                     & \textbf{Timestamp}             & \multicolumn{1}{l}{\textbf{Event Type}} \\ \hline
		\multicolumn{1}{|c|}{945} & \multicolumn{1}{|c|}{$e_1$} &
		\multicolumn{1}{c|}{a} & \multicolumn{1}{c|}{\begin{tabular}[c]{@{}c@{}} 05-12-2011\end{tabular}}                                                                                 & \multicolumn{1}{c|}{!}                    \\ \hline
		\multicolumn{1}{|c|}{945} & \multicolumn{1}{|c|}{$e_2$} &
		\multicolumn{1}{c|}{\{b, c\}} & \multicolumn{1}{c|}{\begin{tabular}[c]{@{}c@{}}07-12-2011\end{tabular}}                                                                         &  \multicolumn{1}{c|}{!}                    \\ \hline
		\multicolumn{1}{|c|}{945} & \multicolumn{1}{|c|}{$e_3$} &
		\multicolumn{1}{c|}{d}        & \multicolumn{1}{c|}{[06-12-2011, 10-12-2011]} &  \multicolumn{1}{c|}{!}                    \\ \hline
		\multicolumn{1}{|c|}{945} & \multicolumn{1}{|c|}{$e_4$} &
		\multicolumn{1}{c|}{\{a, c\}} & \multicolumn{1}{c|}{\begin{tabular}[c]{@{}c@{}}09-12-2011\end{tabular}}                                                                         &  \multicolumn{1}{c|}{!}                    \\ \hline
		\multicolumn{1}{|c|}{945} & \multicolumn{1}{|c|}{$e_5$} &
		\multicolumn{1}{c|}{e}        & \multicolumn{1}{c|}{\begin{tabular}[c]{@{}c@{}}11-12-2011\end{tabular}}                                                                         &  \multicolumn{1}{c|}{?}                    \\ \hline
	\end{tabular}
\end{table}

Existing process mining tools do not support uncertain data. Therefore, novel techniques to manage and analyze it are needed. \emph{Uncertain Directly-Follows Graphs} (UDFGs) allow representing directly-follows relationships in an event log under conditions of uncertainty in the data. This leads to the discovery of models of uncertain logs through methods based on directly-follows relationships such as the Inductive miner~\cite{pegoraro2019discovering}.

\begin{figure}
	\centering
	\begin{minipage}[t]{.45\textwidth}
		\centering
		\includegraphics[width=\linewidth, keepaspectratio]{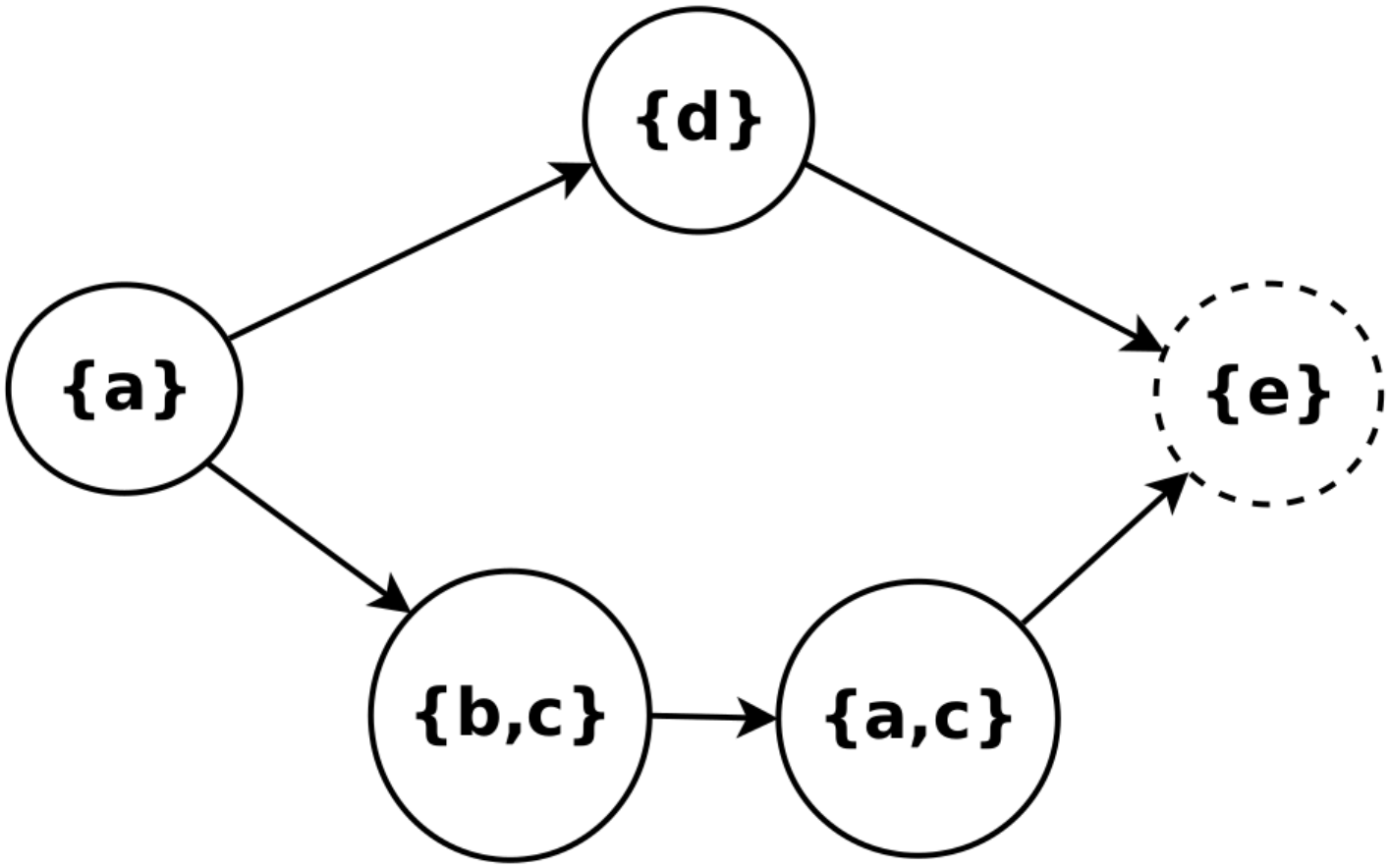}
		\captionsetup{width=.9\linewidth}
		\caption{The behavior graph of the uncertain trace given in Table~\ref{table:uncertaintrace}. Each vertex represents an uncertain event and is labeled with the possible activity label of the event. The dashed circle represents an indeterminate event (may or may not have happened).}
		\label{fig:bg}
	\end{minipage}%
	\quad
	\begin{minipage}[t]{.45\textwidth}
		\centering
		\includegraphics[width=\linewidth, keepaspectratio]{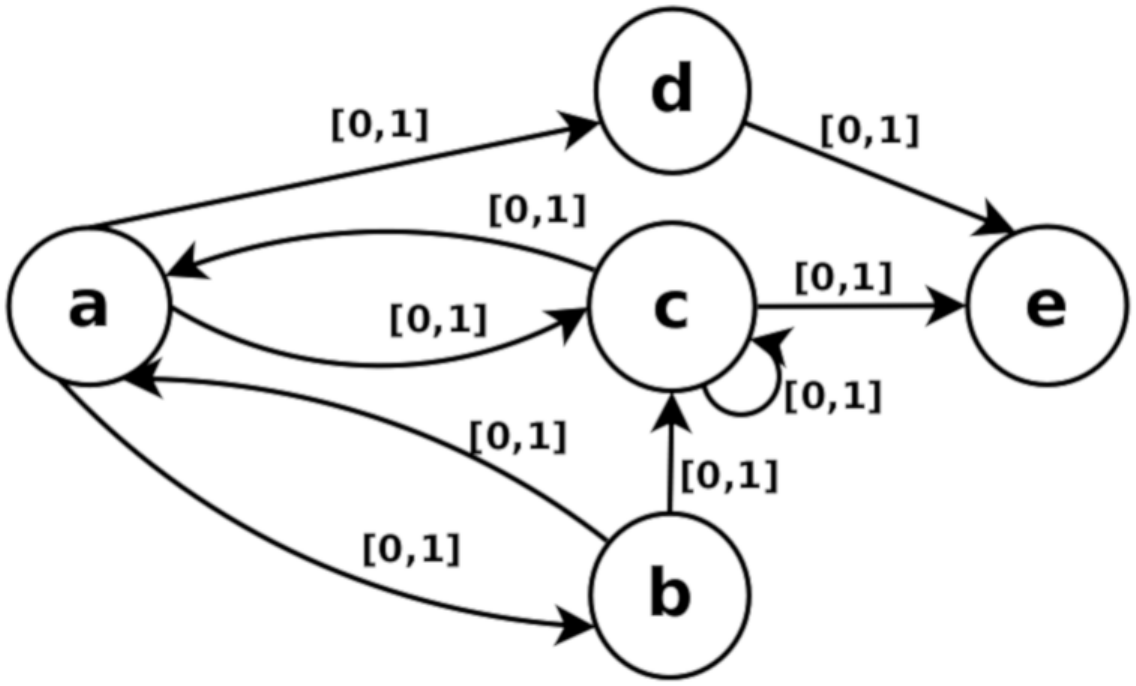}
		\captionsetup{width=.9\linewidth}
		\caption{The UDFG computed based on the behavior graph in Figure~\ref{fig:bg}. The arcs are labeled with the minimum and maximum number of directly-follows relationship observable in the corresponding trace. Here, every relationship can occur in the trace once, or not occur at all.}
		\label{fig:udfg}
	\end{minipage}
\end{figure}

An intermediate step necessary to compute UDFGs is to construct the \emph{behavior graph} of the traces in the uncertain log. A behavior graph represents in a graphical manner the time and precedence relationships among certain and uncertain events in an uncertain trace. Figures~\ref{fig:bg} and~\ref{fig:udfg} show, respectively, the behavior graph of the trace in Table~\ref{table:uncertaintrace} and the UDFG representing the relationship between activities in the same trace. Uncertain timestamps are the most critical source of uncertain behavior in a process trace: for instance, if $n$ events have uncertain timestamps such that their order is unknown, the possible configurations for the control-flow of the trace are the $n!$ permutations of the events.

The construction of behavior graphs for uncertain traces is the basis of both conformance checking and process discovery on uncertain event data. It is, thus, important to be able to build the behavior graph of any given uncertain trace in a quick and efficient manner. Constructing a behavior graph is the most computationally expensive step towards producing a process model (e.g., a Petri net using the approach in~\cite{pegoraro2019mining}). In this paper, we present a novel algorithm for behavior graph construction which runs in quadratic time complexity, therefore allowing a significant speedup for the operations of conformance checking and process discovery for uncertain event logs. We will prove the correctness of the new algorithm, as well as show the improvement in performance both theoretically, via asymptotic complexity analysis, and practically, with experiments on a number of uncertain event logs comparing computing times of the baseline method against the novel construction algorithm. The algorithms have been implemented in the context of the PROVED (\emph{PRocess mining OVer uncErtain Data}) library\footnote{\url{https://github.com/proved-py/proved-core/tree/Efficient_Construction_of_Behavior_Graphs_for_Uncertain_Event_Data}}, based on the PM4Py framework~\cite{berti2019process}.

The reminder of the paper is structured as follows. Section~\ref{sec:related} explores recent related works in the context of uncertain event data. Section~\ref{sec:preliminaries} provides formal definitions and describes the baseline method for our research. Section~\ref{sec:bg} illustrates a novel and more efficient method to construct a behavior graph of an uncertain trace. Section~\ref{sec:proofs} presents the analysis of asymptotic complexity for both the baseline and the novel method. Section~\ref{sec:experiments} shows the results of experiments on both synthetic and real-life uncertain event logs comparing the efficiency of both methods to compute behavior graphs. Section~\ref{sec:conclusions} comments on the results of the experiments and concludes the paper.

\section{Related Work}\label{sec:related}
Research concerning the topic of process mining over uncertain event data is very recent. The work that introduced the concept of uncertainty in process mining, together with a taxonomy of the various kinds of uncertainty, specifically showed that if a trace displays uncertain attributes, it contains behavior, which can be appropriately expressed through process models -- namely, behavior graphs and behavior traces~\cite{pegoraro2019mining}. As opposed to classic process mining, where we have a clear cut between data and model and between the static behavior of data and the dynamic behavior of models, the distinction between data and models becomes blurry in presence of uncertainty, because of the variety in behavior that affects the data. Expressing traces through models is utilized in~\cite{pegoraro2019mining} for the calculation of upper and lower bounds for conformance scores of uncertain traces against classic reference models. A second application for behavior graphs in the domain of process mining over uncertain event data is given in~\cite{pegoraro2019discovering}. Behavior graphs of uncertain traces are employed to count the number of possible directly-follows relationships between uncertain events, with the objective of automatically discovering process models from uncertain event data. The formulation used in this and previous works on uncertainty in process mining shares similarities with temporal extensions of fuzzy logic e.g.~\cite{dutta1988event}; however, unlike fuzzy temporal logic, our framework is suited to compactly represent the control-flow dimension of uncertain event data as Petri nets, a graphical model capable of simulation.

Behavior graphs are Directed Acyclic Graphs (DAGs), which are commonly used throughout many fields of science to represent with a graph-like model time information, precedence relationships, partial orders, or dependencies. They are successfully employed in compiler design~\cite{aho2007compilers}, circular dependency analysis in software~\cite{al2014shape}, probabilistic graphical models~\cite{bayes1763lii} and dynamic graphs analytics~\cite{mariappan2019graphbolt}.

\section{Preliminaries}\label{sec:preliminaries}
Let us introduce some basic notations and concepts, partially from~\cite{van2016process}:

\begin{definition}[Power Set]
	The power set of a set $A$ is the set of all possible subsets of $A$, and is denoted with $\mathcal{P}(A)$. $\mathcal{P}_{NE}(A)$ denotes the set of all the non-empty subsets of $A$: $\mathcal{P}_{NE}(A) = \mathcal{P}(A)\setminus\{\emptyset\}$.
\end{definition}

\begin{definition}[Sequence]
	Given a set $X$, a finite \emph{sequence} over $X$ of length $n$ is a function $s \in X^* : \{1, \dots, n\} \rightarrow X$, and is written as $s = \langle s_1, s_2, \dots, s_n\rangle$. For any sequence $s$ we define $|s| = n$, $s[i] = s_i$, $x \in s \iff x \in \{s_1, s_2, \dots, s_n\}$ and $s \oplus s_0 = \langle s_1, s_2, \dots, s_n, s_0 \rangle$.
\end{definition}

\begin{definition}[Directed Graph]
	A \emph{directed graph} $G = (V, E)$ is a set of \emph{vertices} $V$ and a set of directed \emph{edges} $E \subseteq V \times V$. We denote with $\mathcal{U}_G$ the universe of such directed graphs.
\end{definition}

\begin{definition}[Path]
	A \emph{path} over a graph $G = (V, E)$ is a sequence of vertices $p = \langle v_1, v_2, \dots v_n \rangle$ with $v_1, \dots, v_n \in V$ and $\forall_{1 \leq i \leq n-1} (v_i, v_{i+1}) \in E$. $P_G(v, w)$ denotes the set of all paths connecting $v$ and $w$ in $G$. A vertex $w \in V$ is \emph{reachable} from $v \in V$ if there is at least one path connecting them: $|P_G(v, w)|>0$.
\end{definition}

\begin{definition}[Transitive Reduction]
	The \emph{transitive reduction} of a graph $G = (V, E)$ is a graph $\rho(G) = (V, E')$ with the same reachability between vertices and a minimal number of edges. $E' \subseteq E$ is a smallest set of edges such that $|P_{\rho(G)}(v, w)|>0 \implies |P_G(v, w)|>0$ for any $v, w \in V$. The transitive reduction of a directed acyclic graph is unique.
\end{definition}

This paper analyzes \emph{uncertain event logs}. These event logs contain uncertainty information explicitly associated with event data. A taxonomy of different kinds of uncertainty and uncertain event logs has been presented in~\cite{pegoraro2019mining}; we will refer to the notion of \emph{simple uncertainty}, which includes uncertainty without probabilistic information on the control-flow perspective: activities, timestamps, and indeterminate events. Event $e_4$ has been recorded with two possible activity labels ($a$ or $c$). This is an example of strong uncertainty on activities. Some events, e.g. $e_3$, do not have a precise timestamp, but have a time interval in which the event could have happened has been recorded: in some cases, this causes the loss of the precise order of events (e.g. $e_3$ and $e_4$). These are examples of strong uncertainty on timestamps. As shown by the ``?'' symbol, $e_5$ is an indeterminate event: it has been recorded, but it is not guaranteed to have happened. Conversely, the ``!'' symbol indicates that the event has been recorded in a correct way, i.e. it certainly occurred in reality (e.g. the event $e_1$).

\begin{definition}[Universes]
	Let $\mathcal{U}_I$ be the set of all the \emph{event identifiers}. Let $\mathcal{U}_C$ be the set of all \emph{case ID identifiers}. Let $\mathcal{U}_A$ be the set of all the \emph{activity identifiers}. Let $\mathcal{U}_T$ be the totally ordered set of all the \emph{timestamp identifiers}. Let $\mathcal{U}_O = \{!, ?\}$, where the ``!'' symbol denotes \emph{determinate events}, and the ``?'' symbol denotes \emph{indeterminate events}.
\end{definition}

\begin{definition}[Simple uncertain events]
	$e = (e_i, A, t_{\text{min}}, t_{\text{max}}, o)$ is a simple uncertain event, where $e_i \in \mathcal{U}_E$ is its event identifier, $A \subseteq \mathcal{U}_A$ is the set of possible activity labels for $e$, $t_{\text{min}}$ and $t_{\text{max}}$ are the lower and upper bounds for the value of its timestamp, and $o$ indicates if is is an indeterminate event. Let $\mathcal{U}_E = (\mathcal{U}_I \times \mathcal{P}_{NE}(\mathcal{U}_A) \times \mathcal{U}_T \times \mathcal{U}_T \times \mathcal{U}_O)$ be the set of all simple uncertain events. Over the uncertain event $e$ we define the projection functions $\pi_{t_{\text{min}}}(e) = t_{\text{min}}$ and $\pi_{t_{\text{max}}}(e) = t_{\text{max}}$.
\end{definition}

\begin{definition}[Simple uncertain traces and logs]
	$\sigma \subseteq \mathcal{U}_E$ is a \emph{simple uncertain trace} if for any $(e_i, A, t_{\text{min}}, t_{\text{max}}, o) \in \sigma$, $t_{\text{min}} < t_{\text{max}}$ and all the event identifiers are unique. $\mathcal{T}_U$ denotes the universe of simple uncertain traces. $L \subseteq \mathcal{T}_U$ is a \emph{simple uncertain log} if all the event identifiers in the log are unique. 
\end{definition}

A necessary step to allow for analysis of simple uncertain traces is to obtain their \emph{behavior graph}. A behavior graph is a directed acyclic graph that synthesizes the information regarding the uncertainty on timestamps contained in the trace.

\begin{definition}[Behavior Graph]\label{def:bg}
	Let $\sigma \in \mathcal{T}_U$ be a simple uncertain trace. A behavior graph $\beta \colon \mathcal{T}_U \to \mathcal{U}_G$ is the transitive reduction of a directed graph $\rho(G)$, where $G = (V, E) \in \mathcal{U}_G$ is defined as:
	\begin{itemize}
		\item $V = \{e \in \sigma \}$
		\item $E = \{(v, w) \mid v, w \in V \wedge \pi_{t_{\text{max}}}(v) < \pi_{t_{\text{min}}}(w)\}$
	\end{itemize}
\end{definition}

The semantics of a behavior graph can effectively convey time and order information regarding the time relationship of the events in the corresponding uncertain trace in a compact manner. For a behavior graph $\beta(\sigma) = (V, E)$ and two events $e_1 \in \sigma$, $e_2 \in \sigma$, $(e_1, e_2) \in E$ if and only if $e_1$ is immediately followed by $e_2$ for some possible values of the timestamps for the events in the trace. A consequence is that if some events in the graph are pairwise unreachable, they might have happened in any order.

Definition~\ref{def:bg} is clear and meaningful from a theoretical standpoint. It accurately describes a behavior graph and the semantics of its components. While useful to understand the purpose of behavior graphs, building them from process traces following this definition -- that is, employing the transitive reduction -- is slow and inefficient. This hinders the analysis of larger logs. It is possible, however, to obtain behavior graphs from traces in a quicker way.

\section{Efficient Construction of Behavior Graphs}\label{sec:bg}
The procedure to efficiently build a behavior graph from an uncertain trace is described in Algorithm~\ref{alg:newbg}. For ease of notation, the algorithm textually indicates some conditions on the timestamp of an event. The keyword \texttt{continue} brings the execution flow to the next iteration of the loop in line 16, while the keyword \texttt{break} stops the execution of the inner loop and brings the execution flow on line 30. A certain event $e$ is associated with one specific timestamp which we refer to as \emph{certain timestamp}. Furthermore, an uncertain event e is associated with a time interval which is determined by two values: minimum and maximum timestamp of that event. An event $e$ has a certain timestamp if and only if $\pi_{t_{\text{min}}}(e) = \pi_{t_{\text{max}}}(e)$. A timestamp $t$ is the minimum timestamp of the event $e$ if and only if $t = \pi_{t_{\text{min}}}(e) \neq \pi_{t_{\text{max}}}(e)$. A timestamp $t$ is the maximum timestamp of the event $e$ if and only if $t = \pi_{t_{\text{max}}}(e) \neq \pi_{t_{\text{min}}}(e)$.

\begin{algorithm}[]
	\caption{Efficient construction of the behavior graph}
	\label{alg:newbg}
	\SetKwInOut{Input}{Input~}
	\SetKwInOut{Output}{Output~}
	\Input{~The uncertain trace $\sigma$.}
	\Output{~The behavior graph $\beta(\sigma) = (V, E)$.}
	
	$V \gets \{e \in \sigma \}$ \tcp*{Set of vertices of the behavior graph}
	
	$E \gets \{\}$ \tcp*{Set of edges of the behavior graph}
	
	$\mathcal{L} \gets \langle \rangle$ \tcp*{List of timestamps and events}
	
	\For{$e \in \sigma$}{
		\If{$e$ has a certain timestamp}{
			$\mathcal{L} \gets \mathcal{L} \oplus (\pi_{t_{\text{min}}}(e), e)$
		}
		\Else{
			$\mathcal{L} \gets \mathcal{L} \oplus (\pi_{t_{\text{min}}}(e), e)$
			
			$\mathcal{L} \gets \mathcal{L} \oplus (\pi_{t_{\text{max}}}(e), e)$
		}
	}
	sort the elements $(t, e) \in \mathcal{L}$ based on the timestamps
	
	$i \gets 1$
	
	\While{$i < |\mathcal{L}| - 1$}{
		$(t, e) \gets \mathcal{L}[i]$
		
		\If{$e$ has a certain timestamp or $t$ is the maximum timestamp of $e$}{
			$j \gets i + 1$
			
			\While{$j < |\mathcal{L}|$}{
				$(t', e') \gets \mathcal{L}[j]$
				
				\If {$t'$ is the minimum timestamp of $e'$}{
					$E \gets E \cup \{(e, e')\}$
					
					\Continue
				}
				\If {$e'$ has a certain timestamp}{
					$E \gets E \cup \{(e, e')\}$
					
					\Break
				}
				\If {$t'$ is the maximum timestamp of $e'$}{
					\If {$(e, e') \notin E$}{
						
						\Continue
					}
					\Else{
						\Break
					}
				}
			$j \gets j + 1$
			}
		}
		$i \gets i + 1$
	}
	\Return $(V, E)$
\end{algorithm}

\begin{table}[]
	\caption{Running example for the construction of the behavior graph.}
	\label{table:running}
	\centering
	\begin{tabular}{ccccc}
		\textbf{Case ID} & \textbf{Event ID}        & \textbf{Activity}                                                                                                     & \textbf{Timestamp}             & \multicolumn{1}{l}{\textbf{Event Type}} \\ \hline
		\multicolumn{1}{|c|}{872} & \multicolumn{1}{|c|}{$e_1$} &
		\multicolumn{1}{c|}{a} & \multicolumn{1}{c|}{\begin{tabular}[c]{@{}c@{}} 05-12-2011\end{tabular}}                                                                                 & \multicolumn{1}{c|}{!}                    \\ \hline
		\multicolumn{1}{|c|}{872} & \multicolumn{1}{|c|}{$e_2$} &
		\multicolumn{1}{c|}{b} & \multicolumn{1}{c|}{\begin{tabular}[c]{@{}c@{}}07-12-2011\end{tabular}}                                                                         &  \multicolumn{1}{c|}{!}                    \\ \hline
		\multicolumn{1}{|c|}{872} & \multicolumn{1}{|c|}{$e_3$} &
		\multicolumn{1}{c|}{c}        & \multicolumn{1}{c|}{[06-12-2011, 10-12-2011]} &  \multicolumn{1}{c|}{!}                    \\ \hline
		\multicolumn{1}{|c|}{872} & \multicolumn{1}{|c|}{$e_4$} &
		\multicolumn{1}{c|}{d} & \multicolumn{1}{c|}{\begin{tabular}[c]{@{}c@{}}[08-12-2011, 11-12-2011]\end{tabular}}                                                                         &  \multicolumn{1}{c|}{!}                    \\ \hline
		\multicolumn{1}{|c|}{872} & \multicolumn{1}{|c|}{$e_5$} &
		\multicolumn{1}{c|}{e}        & \multicolumn{1}{c|}{\begin{tabular}[c]{@{}c@{}}09-12-2011 \end{tabular}}                                                                         &  \multicolumn{1}{c|}{!}                    \\ \hline
		\multicolumn{1}{|c|}{872} & \multicolumn{1}{|c|}{$e_6$} &
		\multicolumn{1}{c|}{f}        & \multicolumn{1}{c|}{[12-12-2011, 13-12-2011]}                                                                         &  \multicolumn{1}{c|}{!}                    \\ \hline
	\end{tabular}
\end{table}

We will consider here the application of Algorithm~\ref{alg:newbg} on a running example, the trace shown in Table~\ref{table:running}. Notice that none of the events in the running example display uncertainty on activity labels or are indeterminate: this is due to the fact that the topology of a behavior graph only depends on the (uncertain) timestamps of events.

The concept behind the algorithm is to inspect the time relationship between uncertain events in a more specific way, instead of adding many edges to the graph and then deleting them via transitive reduction. This is achieved by searching the possible successors of each event in a sorted list of timestamps. We then scan the list of timestamps with two nested loops, and we use the inner loop to search for successors of the event selected by the outer loop. It is important to notice that, since the semantics of the behavior graph state that events with overlapping intervals as timestamps should not be connected by a path, we draw outbound edges from an uncertain event only when, scanning the list, we encounter the timestamp at which the event has certainly occurred. This is the reason why outbound edges are not drawn from minimum timestamps (line 14) and inbound edges are not drawn into maximum timestamps (lines 24-28).

\begin{figure}
	\centering
	\begin{minipage}[t]{.5\textwidth}
		\centering
		\includegraphics[width=\linewidth, keepaspectratio]{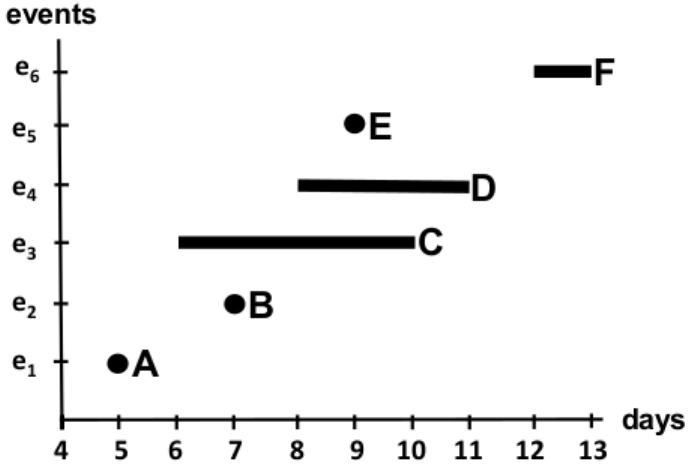}
		\captionsetup{width=.9\linewidth}
		\caption{A diagram visualizing the time perspective of the events in Table~\ref{table:running}.}
		\label{fig:bg_running_table}
	\end{minipage}%
	\begin{minipage}[t]{.4\textwidth}
		\centering
		\includegraphics[width=\linewidth, keepaspectratio]{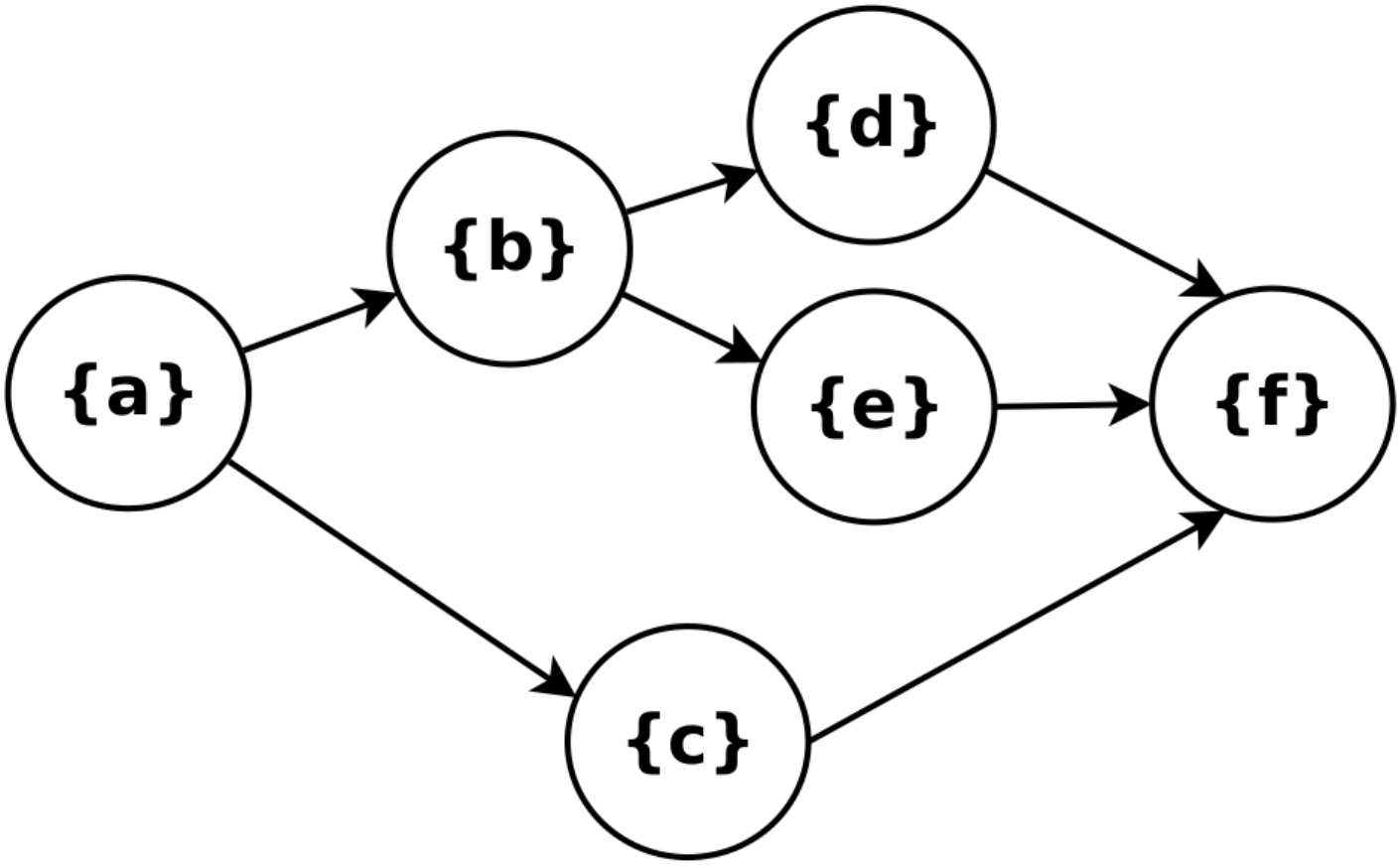}
		\captionsetup{width=.9\linewidth}
		\caption{The behavior graph of the trace in Table~\ref{table:running}.}
		\label{fig:bg_running}
	\end{minipage}
\end{figure}

If, while searching for successors of the event $e$, we encounter the minimum timestamp of the event $e'$, we connect them, since their timestamps do not overlap. The search for successors needs to continue, since it is possible that other events occurred before the maximum timestamp of $e'$ (lines 18-20). This happens for the events $e_1$ and $e_3$ in Table~\ref{table:running}. As shown in Figure~\ref{fig:bg_running_table}, $e_3$ can indeed follow $e_1$, but the undiscovered event $e_2$ is another possible successor for $e_1$. 

If we encounter a certain event $e'$, we connect $e$ with $e'$ and we stop the search. A certain event $e'$ will in fact preclude an edge from $e$ to any event occurring after $e'$ (lines 21-23). The trace in Table~\ref{table:running} shows this situation for events $e_1$ and $e_2$: once connected, nothing that occurs after the timestamp of $e_2$ can be a successor of $e_1$.

If we encounter the maximum timestamp of the event $e'$ (line 24), there are two distinct situations to consider. Case 1: $e$ was not already connected to $e'$. Then, either $e$ is certain and occurred within the timestamp interval of $e'$, or both timestamps of $e$ and $e'$ are uncertain and overlap with each other. In both situations, $e$ should not be connected to $e'$ and the search should continue (lines 25-26). Events $e_3$ and $e_4$ are an example: when the maximum timestamp of $e_4$ is encountered during the search for the successor of $e_3$, the two are not connected, so the search for a viable successor of $e_3$ continues. Case 2: $e$ and $e'$ are already connected. This means that we had already encountered the minimum timestamp of $e'$ during the search for the successors of $e$. Since the whole time interval associated with the timestamp of $e'$ is detected after the occurrence of $e$, there are no further events to consider as successors of $e$ and the search stops (lines 27-28). In the running example, this happens between $e_5$ and $e_6$: when searching for the successors of $e_5$, we first connect it with $e_6$ when we encounter its minimum timestamp; we then encounter its maximum timestamp, so no other successive event can be a successor for $e_5$.

\section{Asymptotic Complexity}\label{sec:proofs}
Definition~\ref{def:bg} provides a baseline method for the construction of the behavior graph consists of two main parts: the creation of the initial graph and its transitive reduction. Let us consider an uncertain trace $\sigma$ of length $n = |\sigma|$ (with $n$ events). Both the initial graph $G = (V, E)$ and the behavior graph $\beta(\sigma)$ have thus $|V| = n$ vertices. The initial graph is created by checking the time relationship between every pair of events; this is equivalent of checking if an edge exists between each pair of vertices of $G$, which is done in $\mathcal{O}(n^2)$ time.

The transitive reduction can be attained through many methods.
Aho et al.~\cite{aho1972transitive} show a method to perform transitive reduction in $\mathcal{O}(n^3)$ time, better suited for dense graphs, and prove that the transitive reduction has the same complexity as the matrix multiplication. The Strassen algorithm~\cite{strassen1969gaussian} can multiply matrices in $\mathcal{O}(n^{2.807355})$ time.
Subsequent improvements have followed suit: the asymptotically fastest algorithm has been described by Le Gall~\cite{le2014powers}. However, these improved algorithms are rarely used in practice, because of the large constant factors in their computing time hidden by the asymptotic notation, as well as very large memory requirements. The Strassen algorithm is useful in practice only for large matrices~\cite{d2005using}, and the Coppersmith-Winograd algorithm and successive improvements require an input so large to be efficient that they are effectively classified as galactic algorithms~\cite{le2012faster}.

In light of these considerations, for the vast majority of event logs the best way to implement the construction of the behavior graph through transitive reduction runs in $\mathcal{O}(n^2) + \mathcal{O}(n^3) = \mathcal{O}(n^3)$ time in the worst-case scenario.

It is straightforward to find the upper bound for complexity of Algorithm~\ref{alg:newbg}. Lines 1-3 and line 11 run in $\mathcal{O}(1)$ time. The worst case scenario is when all events in a trace are uncertain. In that case, lines 4-5 build a list of length $2n$ with a single pass through the events in the trace, and thus run in $\mathcal{O}(n)$. Line 10 sorts the list, running in $\mathcal{O}(2n \log(2n)) = \mathcal{O}(n \log n)$. Lines 11-30 consist of two nested loops over the list, resulting in a $\mathcal{O}((2n)^2) = \mathcal{O}(n^2)$. The total running time for the novel method is then $\mathcal{O}(1) + \mathcal{O}(n) + \mathcal{O}(n \log n) + \mathcal{O}(n^2) = \mathcal{O}(n^2)$ time in the worst-case scenario.

\section{Experiments}\label{sec:experiments}
Both the baseline algorithm~\cite{pegoraro2019mining} and the novel algorithm for the construction of the behavior graph are implemented in Python, in the context of the PROVED project within the PM4Py framework. The experiments are designed to investigate the difference in performances between the two algorithms, and specifically how this difference scales with the increase of the size of the event log, as well as the number of events in the log that have uncertain timestamps.

For each series of experiments, we generate a synthetic event log with $n$ many traces of length $l$ (indicating the number of events in the trace). Uncertainty on timestamps is added to the events in the log. A percentage $p$ of the events in the event log will have an uncertain timestamp, causing it to overlap with adjacent events. Finally, behavior graphs are obtained from all the traces in the event log with either algorithm, while the execution time is measured. All results are shown as an average of 10 runs of the corresponding experiment.

\begin{figure}
	\centering
	\begin{subfigure}{.45\textwidth}
		\centering
		\includegraphics[width=\linewidth, keepaspectratio]{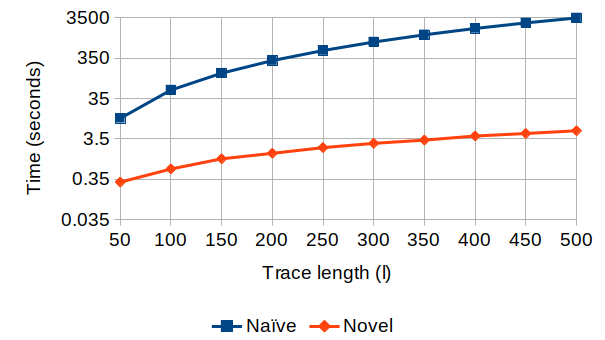}
		\caption{\scriptsize Time in seconds for the creation of the behavior graphs for synthetic logs with $n = 1000$ traces and $p = 0.4$ of uncertain events, with increasing trace length.\normalsize}
		\label{fig:lengths}
	\end{subfigure}%
	\quad
	\begin{subfigure}{.45\textwidth}
		\centering
		\includegraphics[width=\linewidth, keepaspectratio]{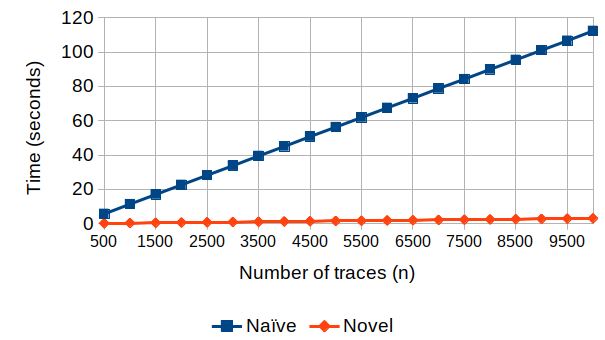}
		\caption{\scriptsize Time in seconds for the creation of the behavior graphs for synthetic logs with traces of length $l = 10$ events and $p = 0.4$ of uncertain events, with increasing number of traces.\normalsize}
		\label{fig:ntraces}
	\end{subfigure}
	\caption{Results of the first and second experiments. The diagrams show the improvement in speed attained by our novel algorithm.}
\end{figure}

In the first experiment, we analyze the effect of the trace length on the overall time required for behavior graph construction. To this end, we generate logs with $n = 1000$ traces of increasing lengths, and added uncertain timestamps to events with $p = 0.4$. The results, presented in Figure~\ref{fig:lengths}, match our expectations: the computing time of the na{\"i}ve algorithm scales much worse than the time of our novel algorithm, due to its cubic asymptotic time complexity. This confirms the findings of the asymptotic time complexity analysis discussed in Section~\ref{sec:proofs}. We can observe order-of-magnitude speedup. At length $l = 500$, the novel algorithm runs in $0.16\%$ of the time needed by the na{\"i}ve algorithm.

The second experiment verifies how the speed of the two algorithms scales with the log dimension in number of traces. We create logs with a trace length of $l = 50$, and a fixed uncertainty percentage of $p = 0.4$. The number of traces scales from $n = 500$ to $n = 10000$. As presented in Figure~\ref{fig:ntraces}, our proposed algorithm outperforms the na{\"i}ve algorithm, showing a relatively smooth behavior exposing a much smaller slope. As expected, the elapsed time to create behavior graphs scales linearly with the number of traces in the event log for both algorithms.

\begin{figure}
	\centering
	\begin{subfigure}{.45\textwidth}
		\centering
		\includegraphics[width=\textwidth, keepaspectratio]{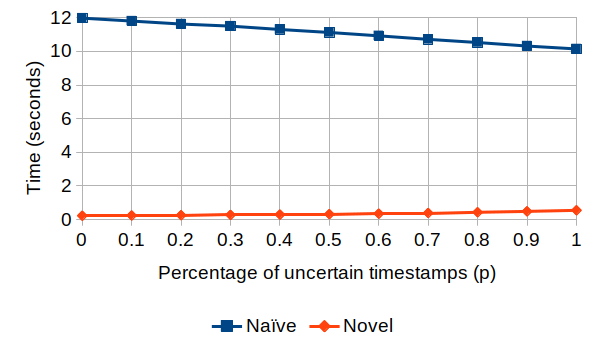}
		\caption{\scriptsize Time in seconds for the creation of the behavior graphs for synthetic logs with $n = 1000$ traces of length $l = 10$ events, with increasing percentages of timestamp uncertainty.\normalsize}
		\label{fig:probs}
	\end{subfigure}%
	\quad
	\begin{subfigure}{.45\textwidth}
		\centering
		\scriptsize
		\begin{tabular}[b]{cccc}
			\toprule
			Event Log & $p$ & Time (na{\"i}ve) & Time (novel)\\
			\midrule
			\multirow{3}*{Help Desk}
			& $0$ & $1.17$ & $0.15$ \\
			& $0.4$ & $1.11$ & $0.17$ \\
			& $0.8$ & $1.06$ & $0.20$ \\
			
			\midrule
			\multirow{3}*{Road Traffic}
			& $0$ & $31.69$ & $4.09$ \\
			& $0.4$ & $30.73$ & $5.05$ \\
			& $0.8$ & $29.45$ & $5.79$ \\
			
			\midrule
			\multirow{3}*{BPIC 2012}
			& $0$ & $58.25$ & $1.50$ \\
			& $0.4$ & $55.22$ & $2.37$ \\
			& $0.8$ & $51.79$ & $3.33$ \\
			
			\bottomrule
		\end{tabular}
		\normalsize
		\caption{\scriptsize Execution times in seconds for real-life event logs with increasing percentages of timestamp uncertainty.\normalsize}
		\label{fig:reallifeprobs}
	\end{subfigure}
	\caption{Effects of different percentages of uncertain timestamps in a trace on the execution time for both algorithms.}
\end{figure}

Finally, the third experiment inspects the difference in execution time for the two algorithms as a function of the percentage of uncertain events in the event log. Keeping the values $n = 1000$ and $l = 50$ constant, we scaled up the percentage $p$ of events with an uncertain timestamp and measured computing time. As presented in Figure~\ref{fig:probs}, the time required for behavior graph construction remains almost constant for our proposed algorithm, while it is decreasing for the na{\"i}ve algorithm. This behavior is expected, and is justified by the fact that a worst-case scenario for the na{\"i}ve algorithm is a trace that has no uncertainty on the timestamp: in that case, the behavior graph is simply a chain of nodes, thus the transitive reduction needs to remove a high number of edges from the graph. Notice, however, that for all possible values of $p$ the novel algorithm runs is faster than the na{\"i}ve algorithm: with $p = 0$, the new algorithm takes $1.91\%$ of the time needed by the baseline, while for $p = 1$ this figure grows to $5.41\%$.

We also compared the elapsed time for behavior graphs construction on real-life event log, where we simulated uncertainty in progressively increasing percentage of events as described for the experiments above. We analyzed three event logs: an event log related to the help desk process of an Italian software company, a log related to the management of road traffic fines in an Italian municipality, and a log from the BPI Challenge 2012 related to a loan application process. The results, shown in Figure~\ref{fig:reallifeprobs}, closely adhere to the findings of the experiments on synthetic uncertain event data.

In summary, the results of the experiments illustrate how the novel algorithm hereby presented outperforms the previous algorithm for constructing the behavior graph on all the parameters in which the problem can scale in dimensions. The third experiment shows that, like the baseline algorithm, our novel method being is essentially impervious to the percentage of events with uncertain timestamps in a trace. While for every combination of parameters we benchmarked the novel algorithm runs in a fraction of time required by the baseline method, the experiments also empirically confirm the improvements in asymptotic time complexity shown through theoretical complexity analysis.

\section{Conclusions}\label{sec:conclusions}
The construction of the behavior graph -- a fundamental structure for the analysis of uncertain data in process mining -- plays a key role as processing step for both process discovery and conformance checking of traces that contain events with timestamp uncertainty, the most critical type of uncertain behavior. In this paper we improve the performance of uncertainty analysis by proposing a novel algorithm that allows for the construction of behavior graphs in quadratic time in the length of the trace. We argued for the correctness of this novel algorithm, the analysis of its asymptotic time complexity, and implemented performance tests for this algorithm. These show the speed improvement in real-world scenarios.

Further research is needed to inspect the capabilities of the novel algorithm. Future work includes extending the asymptotic time complexity analysis presented in this paper with lower bound and average case scenario analysis. Furthermore, behavior graphs are memory-expensive; we plan to address this through a multiset of graphs representation for event logs.

\bibliographystyle{splncs04}
\bibliography{uncertainty_behavior_graph}

\end{document}